\documentclass[12pt,preprint]{aastex}
\newcommand{\omc}{\mbox{$\omega$ Cen~}} 
\newcommand{\omcp}{\mbox{$\omega$ Cen}} 
\newcommand{\rrc}{\mbox{$RR_c\,$}} 
\newcommand{\rrab}{\mbox{$RR_{ab}\,$}} 
\newcommand{\tHB}{\tau_{\mbox{\scriptsize HB}}} 
\newcommand{\PLZk}{PLZ$_K$}

\shorttitle{A Pulsation distance to {\boldmath $\omega$}\,Cen based on 
near-infrared Period-Luminosity relations of RR Lyrae stars} 
\shortauthors{Del Principe et al.}

\begin{document}

%%%%%%%%%%%%%%
% Title Page %
%%%%%%%%%%%%%%

%\slugcomment{\it Accepted to ApJ}

\title{A pulsational distance to {\boldmath $\omega$}\,Centauri based on 
Near-Infrared Period-Luminosity relations of RR Lyrae stars\altaffilmark{1}}

\author{
M. Del Principe\altaffilmark{2},
A. M. Piersimoni\altaffilmark{2},
J. Storm\altaffilmark{3},
F. Caputo\altaffilmark{4},
G. Bono\altaffilmark{4},
P. B. Stetson\altaffilmark{5,11,12},
M. Castellani\altaffilmark{4},
R. Buonanno\altaffilmark{6},
A. Calamida\altaffilmark{6},
C. E. Corsi\altaffilmark{4}
M. Dall'Ora \altaffilmark{7},
I. Ferraro\altaffilmark{4}, 
L. M. Freyhammer\altaffilmark{8},
G. Iannicola\altaffilmark{4}, 
M. Monelli\altaffilmark{9},
M. Nonino\altaffilmark{10},
L. Pulone\altaffilmark{4}, 
V. Ripepi\altaffilmark{7}
}

\altaffiltext{1}{
Based in part on observations collected with the New Technology Telescope 
operated at ESO, La Silla, under programmes: 066D-0557; 068D-0545}

\altaffiltext{2}{INAF-Osservatorio Astronomico di Collurania, via M. Maggini, 
64100 Teramo, Italy; milena@te.astro.it, piersimoni@te.astro.it 
}

\altaffiltext{3}{Astrophysikalisches Institut Potsdam, An der Sternwarte 16, 
D-14482 Potsdam, Germany; jstorm@aip.de 
}

\altaffiltext{4}{
 INAF-Osservatorio Astronomico di Roma, Via Frascati 33, 00040, Monte Porzio 
Catone, Italy; bono@mporzio.astro.it, buonanno@mporzio.astro.it, 
calamida@mporzio.astro.it, caputo@mporzio.astro.it, corsi@mporzio.astro.it, 
ferraro@mporzio.astro.it, giacinto@mporzio.astro.it, m.castellani@mporzio.astro.it,
pulone@mporzio.astro.it}
\altaffiltext{5}{
 Dominion Astrophysical Observatory, Herzberg Institute of Astrophysics,
 National Research Council, 5071 West Saanich Road, Victoria, BC V9E~2E7,
 Canada; Peter.Stetson@nrc-cnrc.gc.ca}
\altaffiltext{6}{
Universita' di Roma Tor Vergata, Via della Ricerca Scientifica 1,
00133 Rome, Italy}
\altaffiltext{7}{INAF - Osservatorio Astronomico di Capodimonte,
 via Moiariello 16, 80131 Napoli; dallora@na.astro.it,  ripepi@na.astro.it
 }
\altaffiltext{8}{Centre for Astrophysics, University of Central Lancashire, 
 Preston PR1 2HE; lmfreyhammer@uclan.ac.uk 
 }
\altaffiltext{9}{IAC - Instituto de Astrofisica de Canarias, Calle Via Lactea,
 E38200 La Laguna, Tenerife, Spain; monelli@iac.es
 }
\altaffiltext{10}{INAF-Osservatorio Astronomico di Trieste, via G.B. Tiepolo 11,
 40131 Trieste, Italy; nonino@oats.inaf.it  
 } 
\altaffiltext{11}{Guest User, Canadian Astronomy Data Centre, which is operated
 by the Herzberg Institute of Astrophysics, National Research Council of Canada.}
\altaffiltext{12}{Guest Investigator of the UK Astronomy Data Centre.}
% % 

%\date{\centering drafted \today\ / Received / Accepted }
% 
% %%%%%%%%%%%%
% Abstract %
%%%%%%%%%%%%
\begin{abstract}
We present new Near-Infrared ($J$,$K$) magnitudes for 114~RR~Lyrae
stars in the globular cluster \omc (NGC~5139) which we combine with
data from the literature to construct a sample of 180 RR Lyrae stars
with $J$ and $K$ mean magnitudes on a common photometric system. This
is presently the largest such sample in any stellar system.  We also
present updated predictions for $J$,$K$-band Period-Luminosity relations
for both fundamental and first-overtone RR Lyrae stars, based on 
synthetic horizontal branch models with metal abundance ranging from 
$Z$=0.0001 to $Z$=0.004. 
By adopting for the \omc variables with measured metal abundances an
$\alpha$-element enhancement of a factor of 3 ($\approx 0.5$ dex) with
respect to iron we find a true distance modulus $\mu_0 = 13.70\pm0.06\pm0.06$
(random and systematic errors respectively), corresponding to a distance
$d=5.5\pm0.03\pm0.03$~Kpc.
Our estimate is in excellent agreement with the 
distance inferred for the eclipsing binary OGLEGC-17, but differ significantly 
from the recent distance estimates based on cluster dynamics and on high 
amplitude $\delta$ Scuti stars.
\end{abstract}

\keywords{globular clusters: $\omega$ Centauri -- stars: evolution --
stars: horizontal branch -- stars: oscillations -- stars: variables:
other}

%%%%%%%%
% Body %
%%%%%%%%

\section{Introduction} \label{introduction}

The Period-Luminosity ($PL$) relation for Classical Cepheids 
and, in the near-infrared, for RR Lyrae stars is one of the 
most important methods to estimate distances and to calibrate 
secondary distance indicators 
(Bono et al. 2001; Freedman et al. 2001; Saha et al. 2001; 
Cassisi et al. 2004, hereinafter C04; Catelan, Pritzl, \& Smith 2004, 
hereinafter CPS04; Gieren et al. 2005). 
Current trigonometric parallaxes from
Hipparcos (Feast \& Catchpole 1997), HST (Benedict et al. 2002), and 
from ground based telescopes (Lane et al. 2000; Kervella et al. 2004) 
have provided a unique possibility to 
calibrate the zero-point of these relations. Unfortunately, the 
number of Galactic calibrators with accurate
trigonometric parallaxes is very limited, and for RR Lyrae stars it is
restricted to a single object, namely RR Lyr itself. Therefore, 
we still lack a sound knowledge of the slope of the different PL relations 
based on purely geometrical methods. To overcome this problem, the astronomical 
community devoted a paramount observational effort to measure the 
pulsation properties of these robust distance indicators.
The ideal laboratories for these studies are the Globular Clusters (GCs) 
and the dwarf spheroidals (dSphs) in the Local Group as the RR Lyrae 
stars in a given system are located at largely the same distance.

It has been demonstrated by Longmore et al. (1990) that cluster RR
Lyrae follow a well-defined $K$-band $PL$ relation but we still lack
a comprehensive empirical investigation of its properties.  For this
reason, during the last few years we undertook a detailed pulsation
and evolutionary investigation of RR Lyrae stars (Bono et al. 2003 and
references therein) and, at the same time, we also started a large
observational project aimed at collecting new Near-Infrared (NIR)
data for RR Lyrae stars in  Galactic (M92: Del Principe et al. 2005)
and Large Magellanic Cloud clusters (Reticulum: Dall'Ora et al. 2004).

Although the theoretical and empirical scenarios have recently seen 
new predictions based on Synthetic Horizontal Branch (SHB) models 
(Cassisi et al. 2004); Catelan, Pritzl, \& Smith 2004) and new data 
(Butler 2003; Borissova et al. 2004; Storm 2004), we still 
lack a detailed NIR photometric analysis of a stellar system which 
includes a large sample of fundamental (F, \rrab) and first-overtone 
(FO, \rrc) RR Lyrae stars (only a dozen Galactic GCs host more than 
50 RR Lyrae, see, e.g., Clement et al. 2001) and which spans 
a broad range in metal abundances. At present,  
the largest NIR data set includes 74 RR Lyrae stars whose metal abundance, 
based on 23 objects, ranges from [Fe/H]$\approx-2$ to [Fe/H]$\approx-0.7$, 
as  observed by Borissova et al. (2004) in the inner regions of the Large 
Magellanic Cloud.  In this context, the GC \omc is a perfect target, since 
it  includes more than 186 RR Lyrae (86 \rrab, 100 \rrc; Kaluzny et al. 2004) 
with a metal content ranging from [Fe/H]$\sim-2$ to [Fe/H]$ \sim-1$ 
(131 objects, Rey et al 2000; 74 objects Sollima et al. 2006).  
Moreover, the \omc distance has been estimated using different distance 
indicators such as the First Overtone Blue Edge (FOBE, Caputo et al. 2002),
the eclipsing binary OGLEGC-17 (Thompson et al.  2001; Kaluzny et al. 2002),
the $V$-band metallicity relation of RR Lyrae stars (Rey et al. 2000; 
Catelan 2005),
and a dynamical analysis by van de Ven et al. (2006) based on proper
motion and radial velocities of a large sample of individual stars.

In this paper, we present new NIR 
photometry for a substantial number of the RR Lyrae stars in \omc and
we provide new distance estimates which we compare with determinations
already present in the literature.

%%%%%%%%%%%%%%%%%%%%%%%%%%%%%%%%%%%%%%%%%%%%%%%%%%%%%%%%%%%%%%%%%%%%%%%
\section{Observations and data reduction} \label{observations}

The present NIR data set includes three different samples of $J$,$K_s$ data 
either collected with the NIR camera SOFI available at the New Technology 
Telescope (NTT, ESO, La Silla), or by Longmore et al. (1990), or available 
from the Two Micron All Sky Survey (2MASS) catalogue\footnote{See
http://www.ipac.caltech.edu/2mass/releases/allsky/}.  

\begin{figure}
\begin{center}
\includegraphics[height=0.45\textheight,width=0.60\textwidth]{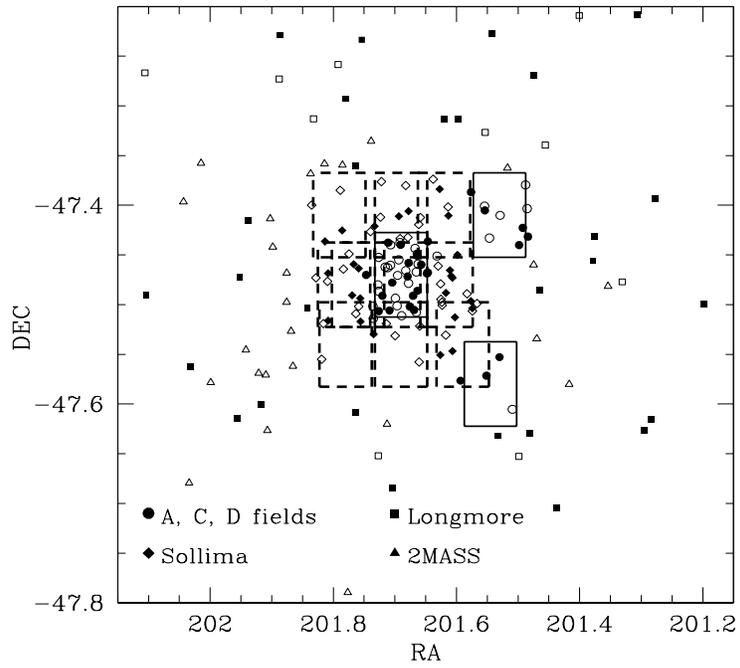}
\caption{Distribution of RR Lyrae stars in the field of \omcp. 
Open and filled symbols show first-overtone and fundamental mode RR Lyrae. 
 Solid and dashed rectangles show the fields observed with SOFI@NTT. 
See text for more details.\label{fig1}  
}
\end{center}
\end{figure}

Near-Infrared $J$ and $K_s$ images of \omc were collected in 
two different runs, namely 6,8 February 2001 (UT) and 3,4,25, 
and 26 February 2002 (UT). The pixel scale of SOFI is $0.292\arcsec$ while the 
field of view is $4.94\arcmin\times4.94\arcmin$. We observed three 
different fields: Field A is centered on the cluster center, 
field C at a distance $\sim10.5\arcmin$ NW away from the center, 
and field D at a distance $\sim10.7\arcmin$ SW away from the 
center (see solid rectangles in Fig. \ref{fig1}). 
We collected 36 $J$- and 49 $K_s$-band images of field A, 
33 $J$- and 52 $K_s$-band images of field C, and 
12 $J$- and 18 $K_s$-band images of field D. 
For $J$ images the exposure time was $t=3\,$s, while for 
$K_s$ images was $t=12\,$s.  
Owing to the crowded nature of the fields the observing time was 
equally divided between observations on \omc and sky observations. 
Together with these data we retrieved from the ESO archive a mosaic 
of nine $J$ and nine $K_s$ images of $\omega$ Cen (Sollima et al. 
2004) which cover an area of $\sim 13'\times13'$ across 
the cluster center (see dashed rectangles in Fig. \ref{fig1}). 
 Note that our Field A overlaps with the central pointing of 
this dataset. 
These images were collected on 13,14 January 2000 (UT), using the same 
experimental equipment. Seven out of the nine fields were observed 
once in $J$,$K_s$ bands, while the other two fields were observed 
twice. The individual exposure times were $54\,$s and $180\,$s.

The pre-processing was performed using standard IRAF procedures. 
To improve the photometric accuracy we did not stack the individual 
images, but determined the (linearly variable) PSF for each frame based
on more than 50 PSF stars uniformly distributed across each image.
Photometry for the entire set of $J$ and $K_s$ images was then 
simultaneously performed with DAOPHOT/ALLFRAME (Stetson 1994)
together with a few $V$- and $I$-band images (Dall'Ora et al. 2004; 
Del Principe et al. 2005). A more detailed analysis concerning the 
reduction strategy will be discussed in a forthcoming paper 
(Del Principe et al. 2006, in preparation).  

\begin{figure}
\begin{center}
\includegraphics[height=0.40\textheight,width=0.55\textwidth]{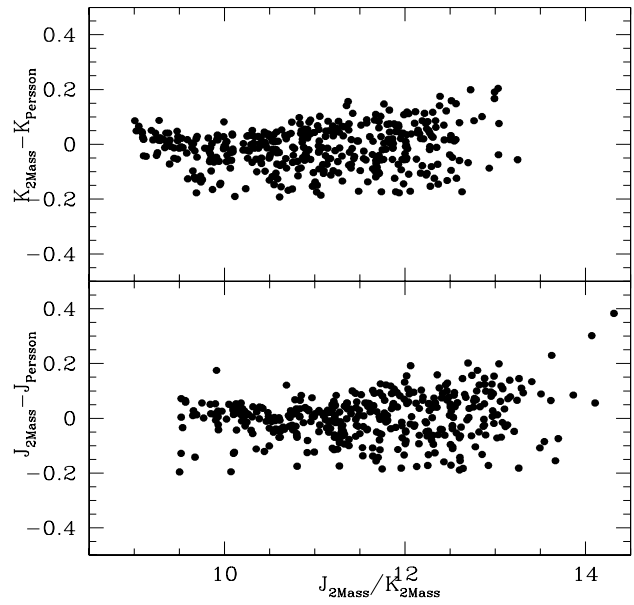}
\caption{Difference in $K$ (top) and $J$ (bottom) magnitudes of Field A 
based on two different absolute zero-point calibrations. 
$J_{Persson}, K_{Persson}$ secondary standards have been calibrated 
using four standard stars measured by Persson et al. (1998), while
$J_{2MASS}, K_{2MASS}$ secondary standards have been calibrated using
$\approx 400$ local 2MASS standard stars. \label{fig2}  
}
\end{center}
\end{figure}

The absolute calibration in the 2MASS system (Cutri et al. 2003) was 
performed using a large set of $\approx 400$ local stars selected from 
the 2MASS point-source catalogue. 
To constrain on a quantitative basis the intrinsic accuracy of both 
absolute and relative zero-points, we performed an independent 
calibration using a set of 4 standard stars (Persson et al. 1998) 
observed during the night of 24 February 2002 at air masses which 
bracket the observations of fields A, C, and D. 
The stars selected were S791-C, S867-V, S273-E, and S870-T.
Fig. \ref{fig2} shows the comparison between secondary 
standards in field A and local 2MASS stars. 
The observed distribution of magnitude differences has a 
standard deviation of $K_{2M} - K_{P} = -0.012\pm0.076$  and 
$J_{2M} - J_{P} =  0.003\pm0.080$, which is a measure of the 
typical difference for an individual star.
Note that the quoted differences agree well, within the errors, with the
transformations provided by Carpenter (2001). As a further independent test, 
we also calibrated the NIR data using standard stars collected during the other 
observing nights, and we found that the accuracy of the absolute zero-points  
is $\sim 0.02\,$~mag.

\begin{figure}
\begin{center}
\includegraphics[height=0.40\textheight,width=0.55\textwidth]{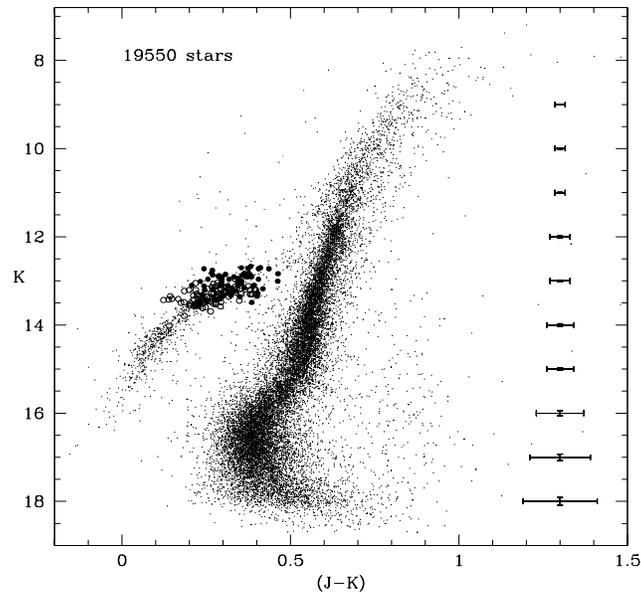}
\caption{NIR Color-Magnitude Diagram of \omcp. The stars plotted in 
this figure are located within $\approx 13'\times13'$
across the cluster center. Open and filled circles mark the sample 
of first-overtone and fundamental RR Lyrae stars for which we 
estimated $J$ and $K$-band magnitudes. Error bars display intrinsic 
photometric errors.  \label{fig3}  
}
\end{center}
\end{figure}

We ended up with a photometric catalogue of $\sim 115,000$ stars with 
limiting magnitudes of $J\approx 20$ and $K_s\approx 19.0$ mag. 
Fig. \ref{fig3} shows the NIR CMD of the stars with a `separation index' , 
$\tt sep$\footnote{The `separation index' estimates the photometric 
contamination due to crowding (Stetson et al. 2003). Current 
$\tt sep$ value corresponds to stars having less than 6\% of their 
observed light contributed by known neighbours.} larger than 3.0 
and an intrinsic photometric accuracy of $\sigma_{J,K} \le 0.08$~mag.

We identified 56 catalogued (OGLEII, Kaluzny et al. 2004) RR Lyrae stars 
(27 \rrab, 29 \rrc) in fields A, C, 
and D  (see circles in Fig. \ref{fig1}) and 58 additional RR Lyrae (25 \rrab, 33 \rrc) 
variables (see diamonds in Fig. \ref{fig1}) in Sollima's fields.
In order to increase the sample of RR Lyrae stars we also included data for 
the 29 RR Lyrae stars (19 \rrab, 10 \rrc; see squares in Fig. \ref{fig1}) observed 
by Longmore et al. (1990). We have transformed their mean $K$-band  data
from the Anglo Australian Observatory NIR system to the 2MASS $K_s$ 
system using the transformations provided by Carpenter (2001). Additionally, 
we cross-correlated the catalogues provided by OGLEII and by Clement et al. (2001)  
with the 2MASS catalogue. In this way we identified 37 more catalogued RR Lyrae 
stars (10 \rrab, 27 \rrc; see triangles in Fig. \ref{fig1}), and therefore we 
ended up with a total sample of 180 RR Lyrae (81\rrab, 99\rrc) for which we 
have at least one $J$ or $K_s$-band measurement.

To obtain the most accurate mean magnitudes for the RR Lyrae,
we fit the individual $K_s$ phase points measured by ALLFRAME with 
a template curve (Jones et al. 1996) and compute on the basis of this
curve the intensity-averaged mean magnitude.
The method requires an accurately known ephemeris of the epoch of 
maximum for each star as well as the amplitude in the optical 
$V$ or $B$ bands.  These parameters are available from
Kaluzny et al. (2004)\footnote{For more details see the home page
of the Cluster AgeS Experiment, http://case.camk.edu.pl} for 167 of the
variables in our sample. For each of the remaining 13 variables, all with 
single epoch 2MASS measurements, the 2MASS measurement was adopted as the 
best estimate of the mean magnitude.

% point I 
The typical accuracy of mean $K-band$ magnitude of RR Lyrae stars either with
a good coverage of the light curve or with single epoch measurements, epoch
of maxima and luminosity amplitudes is better than 0.02 mag.
In order to constrain the typical uncertainty for the remaining thirteen objects
(2MASS sample) with only single epoch measurements, we estimated the difference
between the mean $K-band$ magnitudes of the RR Lyrae stars for which we have a
very good coverage of the light curves (22 \rrab, 22 \rrc) and the mean $K-band$
magnitudes only based on the 2MASS measurements. We found that the individual
differences are smaller than 0.2 mag for \rrab and 0.1 for \rrc.

% point H 
The current RR Lyrae sample includes 16 Blazhko RR Lyrae stars (Kaluzny et al. 2004).
Ten Blazhko RR Lyrae (eight in our sample and two in the Longmore sample) present
a good coverage of the light curves. For four objects (three in the Sollima sample
and one in the 2MASS sample) are available single epoch measurements and accurate
epochs, and therefore the mean magnitude was determined using the template curve.
For only two objects in the 2MASS sample we adopted the single epoch measurements,
since recent epochs are not available. The Blazhko effect shows up as a modulation
in the luminosity amplitude (Stothers 2006, and references therein) that might
affect the accuracy of the mean $K-band$ magnitudes estimated with the template
curve. In order to account for the uncertainty in the luminosity amplitude we
estimated the mean $K-band$ magnitude by increasing/decreasing the amplitude by
a factor of two. The difference in the mean $K-band$ magnitude is of the order
of 0.02 mag. The error for the two objects with single epoch $J$ and $K-band$
magnitudes is at most of the order of 0.15-0.20 mag. 

Unfortunately, no empirical template is available for the $J$-band light 
curves. Instead the mean $J$-band magnitude of RR Lyrae 
stars with well-sampled light curves (56 RR Lyrae stars located in 
fields A,C,D) was determined from a  cubic spline fit to the light
curves. 
% point J 
 The fit was performed using spline under tension (Stellingwerf 1978, and references
therein). This approach presents several advantages when compared with the classical
Fourier fit, in particular for noisy light curves (Del Principe et al. 2006).
For the other objects the $J$-band magnitude is based on a single epoch measurement. 
% point I 
 We found that the typical difference between the mean $J-band$ magnitudes of
the RR Lyrae stars for which we have a very good coverage of the light curves
(22 \rrab, 22  \rrc) and the mean $J-band$ magnitudes only based on single epoch
2MASS measurements is systematically smaller than 0.3~mag for \rrab and 0.15~mag
for \rrc stars. 

Open and filled circles in Fig. \ref{fig3} show the location in the CMD 
of the entire sample of RR Lyrae stars.  This is the largest sample of 
homogeneous NIR magnitudes for RR Lyrae stars in a stellar system ever 
collected.  
 
%%%%%%%%%%%%%%%%%%%%%%%%%%%%%%%%%%%%%%%%%%%%%%%%%%%%%%%%%%%%%%%%%%%%%%%%%%%%%
\section{Distance determinations}\label{Distance_02}

To determine the distance to \omc on the basis of the observed mean
dereddened $K_0$ and $J_0$ magnitudes we need to adopt a relation between
observational quantities and the absolute magnitude.  From the theoretical
point of view, these relations have been recently investigated (e.g., Bono et
al. 2001, 2003; CPS04; C04) on the basis of Synthetic
Horizontal-Branch (SHB) models with promising results in good agreement
with empirical calibrations. These model grids can be parametrized in
different ways, the simplest relation being the $\log P - M_K$ relation
and the more complicated relations taking into account either the global 
metallicity $Z$ (e.g. Bono et al. 2001, 2003) or the Lee (1990) Horizontal Branch
type $\tHB$\footnote{ $\tHB$ is defined as the ratio $\tHB = (B-R)/(B+V+R)$, 
where $B,V,R$ refers to the number of blue, variable, and red HB stars,
respectively, thus resulting in $\tHB=-1$ for HBs with only red stars
and $\tHB=1$ for HBs with only blue stars.} (e.g., C04, and CPS04).

Bono et al. (2003) showed that metallicity has a significant impact on
the $K$-band magnitude of RR Lyrae stars. 
% point K 
 C04 showed that using the HB type as a parameter, at fixed metal 
abundance, also results in tight relations in both $J$ and $K$. 
The latter approach is only valid for globular clusters
where the HB type can be determined, and it is not directly applicable
to field stars. \omc is a peculiar object as it is morphologically
a globular cluster but the chemical composition and probably the ages 
of the stars differ significantly, making the population more similar 
to that of a dwarf galaxy. Consequently, the HB type is not well defined 
a priori, although the color-magnitude diagram presented in Fig. \ref{fig3}
shows that there are very few HB stars redder than the RR Lyrae
stars. This feature suggests that all the sub-populations which 
contribute to the HB must have a collective HB type close to 
1\footnote{Castellani et al. (2006, in preparation) find 
$\tHB=0.94$ on average for \omcp.}.

In order to determine accurate distance estimates for both F and FO 
pulsators, we computed a new set of SHB models with selected metal
abundances ($Z$=0.0001, 0.0003,
0.001, 0.004) and a helium content of $Y=0.245$.  We did
this by adopting the detailed set of HB models recently constructed by
Pietrinferni et al. (2006) and the procedure already discussed by C04.
The pulsation models discussed by Di Criscienzo et al. (2004) were 
used to fix the boundaries of the RR Lyrae
instability strip and to evaluate the periods of the pulsators.  On this
basis, evolutionary and pulsation predictions were used to constrain
the stellar distribution along the HB, and to compute 
% point M 
for each individual simulation the HB-type $\tHB$.
% point N new - 2 
The SHB models were constructed by accounting for the hysteresis effect
(see, e.g., Bono et al. 2005) and by neglecting the pre-ZAHB evolutionary 
phases (see, e.g., Piersanti, Tornambe', Castellani 2004).

Theoretical predictions were transformed into the Bessell \& Brett (1988) NIR  
photometric system  using the bolometric corrections and color-temperature 
relations provided by Pietrinferni et al. (2004). They were then transformed into 
the 2MASS system using the transformations provided by Carpenter (2001) and eventually, 
for each given SHB, the predicted Period-Luminosity relations, as given in the form 
$M_{J,K}=a_{J,K}+b_{J,K}\log P+c_{J,K}\tHB$\footnote{Note that in the following we  
will use $K$-band notation for $K_s$(2MASS)-band.}, were derived. The coefficients 
and the relative errors of the least-squares solutions are listed in 
Table \ref{tab.logPM}.
% point N 
 The zero-points and the coefficients of the above PLZ relations further 
support the use of independent relations for F and FO pulsators. As a matter of 
fact, the slopes and the zero-points between the two different groups differ at 
least at ten $\sigma$ level for both $J$ and K-band. 
% point N new - 1  
The referee suggested to check the dependence of current F and FO
PL relations on the predicted effective temperature of F blue edge
and FO red edge. We performed a few numerical experiments artificially
increasing/decreasing by 150 K the effective temperature of FO red edge
and of F blue edge. We found that the change in the effective temperature
of the F blue edge affects the coefficients of $J$ and $K-$band FO PL
relations by $\pm 0.02-0.03$ ($a$) and by $\pm 0.02-0.05$ ($b$), respectively.
The coefficients of $J$ and $K-$band F PL relations change by $\pm 0.02$ ($a$)
and by $\pm 0.02-0.04$ ($b$), respectively. The change in the effective
temperature of the FO red edge affects the $a,b$ coefficients of F and
FO relations on average by $\pm 0.01$. The impact of these changes on the
coefficient of the HB type ($c$) is marginal. Note that the adopted
uncertainty on the location of FO red edge and F blue edge is supported
by the comparison with RR Lyrae stars in GCs (Di Criscienzo et al. 2004).
These findings further support the approach to derive different
PL relations for F and FO pulsators. It is also noteworthy, that the use
of fundamentalized FO periods causes an increase of a factor of two in
the uncertainty of their zero points and slopes.

As a whole, these relations agree quite well with those   
already presented by C04 (see their Tables 6 and 8). In particular, 
{\em i)} the slopes of the $M_{J,K}-\log P$ relations are, at fixed 
metal content, independent of the HB-type, while they become slightly 
steeper with increasing $Z$.  
{\em ii)} The zero-points of the $M_{J,K}-\log P$ relations become, 
at fixed metal content, brighter when moving from blue to red 
HB populations. This dependence significantly decreases with 
increasing $Z$. 
{\em iii)} The zero-points of the $M_{J,K}-\log P$ relations, 
independently of the HB-type, become fainter with increasing $Z$.
An even more striking agreement is found with the relations computed by
CPS04. For the relevant mean metallicity
model, $Z=0.001$, and HB type, $\tHB=0.934$, they find for their combined
sample of fundamental mode and first-overtone pulsators $M_K = -2.388
\log P_F -1.133$ whereas we find for the fundamental mode pulsators $M_K
= -2.37\log P_F - 1.13$ and for the first-overtone pulsators $M_K =
-2.41\log P_{FO}- 1.20$.
% point Z
The referee noted that, according to CPS04, the coefficients $a$ and $b$ of the
PL relations present some dependence on both the HB type and the metal abundance.
Unfortunately, a global comparison is hampered by the different approach adopted
to derive the PL relations, i.e. multilinear least-squares vs. third order polynomials
(see their equations. 1 and 2 and their Table 9). Moreover, the metallicity of current
SHB models range from Z=0.0001 to Z=0.004, while those from CPS04 range from Z=0.0005
to Z=0.006.

The individual reddening values of \rrab stars have been empirically estimated 
using a Period-Color($V-K$)-Amplitude($A_V$) relation (Piersimoni et al. 2002, 
2006, in preparation), while for \rrc stars and RR Lyrae affected by amplitude 
modulation we used a reddening map based on 
{\em u,v,b,y} Str\"omgren photometry (Calamida et al. 2005). Accurate mean 
$V$-band magnitudes for RR Lyrae in our sample have been measured either by 
Kaluzny et al. (2004)\footnote{For more details see the home page of the 
Cluster AgeS Experiment, http://case.camk.edu.pl} or by Butler et al. (1978)
for the outermost ones. 
For the remaining objects, we adopted a mean reddening of $E(B-V)=0.11\pm0.02$.
Selective absorptions have been estimated using the reddening law from 
Cardelli et al. (1989), that is $A_V/E(B-V)=3.1$, $A_J/E(B-V)=0.868$, 
and $A_K/E(B-V)=0.341$. 

\begin{figure}
\begin{center}
\includegraphics[height=0.65\textheight,width=0.70\textwidth]{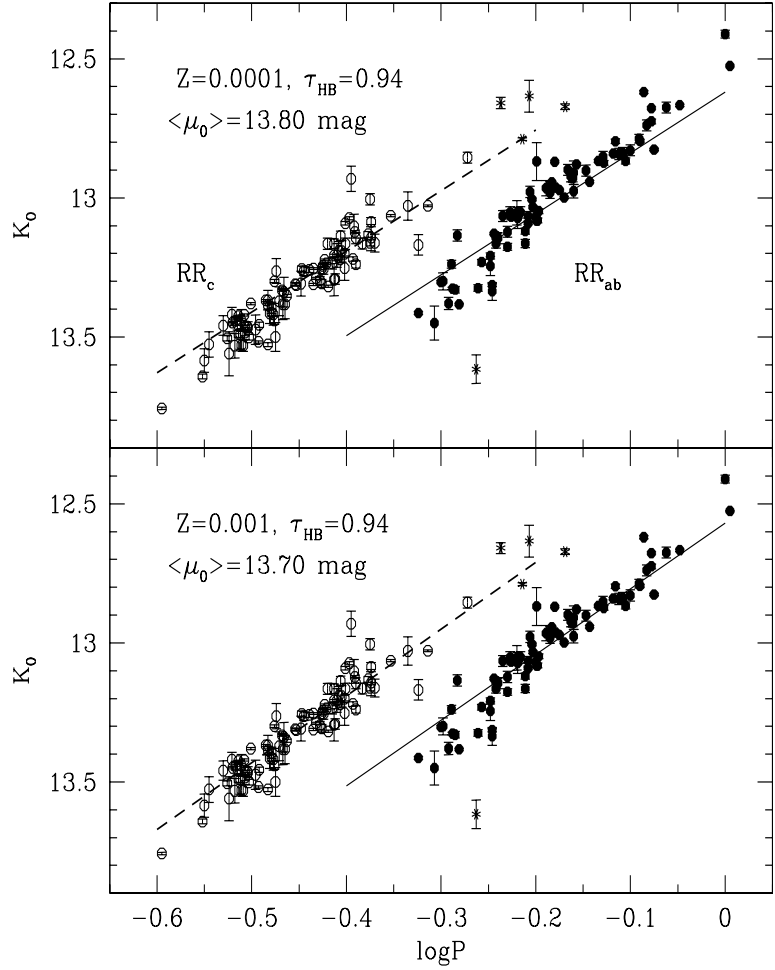}
\caption{Dereddened $K$ magnitudes of RR Lyrae stars in \omc versus period.
Dashed and solid lines display the predicted Period-Luminosity relations
for F (filled circles) and FO (open circles) pulsators, respectively. The 
adopted metal content and HB-type are labeled. The error bars show intrinsic 
errors in the measured magnitudes.\label{fig4}  
}
\end{center}
\end{figure}

Data plotted in Fig. \ref{fig4} show the fit between the dereddened $K$ magnitudes 
and predicted PL relations at fixed metal abundance and for $\tau_{HB}= 0.94$. 
In this figure, the asterisks mark the overluminous \rrab stars which are 
either blended with a fainter close companion 
(V32, V118, V139, and V150) or have a controversial mode 
identification (V84), while the open triangle refers to V182 which, 
together with three other (V168, V181, V183) underluminous variables with  
$K_0\ge$14 mag, presents peculiar $J-K$ colors. By excluding these variables, 
we estimate a true distance of $\langle \mu_0 \rangle=13.80\pm 0.10$ mag and
$13.70\pm 0.10$ mag with $Z$=0.0001 and $Z$=0.001, respectively, 
clearly indicating that individual metallicity estimates are required to 
improve the distance determination.    
 
Conveniently enough, individual metal abundances on the scale of Zinn \& West
(1984), as based on the $hk$ index, for a substantial fraction (130) of the 
RR Lyrae plotted in Fig. \ref{fig4} have been measured by Rey et al. (2000).
In a recent investigation based on medium resolution spectroscopy, 
Sollima et al. (2006) found, for 74 RR Lyrae in \omc, no systematic 
difference with the metal abundances based on the $hk$ index.
To properly use the theoretical predictions, the observed [Fe/H] values have been 
transformed into ``global'' $Z$ values by taking into account possible
$\alpha$-element enhancements\footnote{Following Salaris, Chieffi \& Straniero (1993) 
the global metallicity was estimated as 
$\log Z$=[Fe/H]+log$Z_{\odot}+\log (0.362+0.638f)$, 
where $Z_{\odot}$=0.012 (Asplund et al. 2004) and $f$ is the enhancement 
factor of $\alpha$-elements with respect to iron. 
% point S 
 Note that the change in the distance modulus when moving from scaled solar 
to $\alpha$ enhanced predictions also includes current uncertainties in the 
solar metallicity.}. 
Following to Gratton,
Sneden, \& Caretta (2004), who found [$\alpha$/Fe]$\approx 0.45$ 
as a plausible upper limit to the $\alpha$-element enhancement, we adopt 
$f=10^{[\alpha\mbox{\scriptsize /Fe}]}=3$.

\begin{figure}
\begin{center}
\includegraphics[height=0.65\textheight,width=0.70\textwidth]{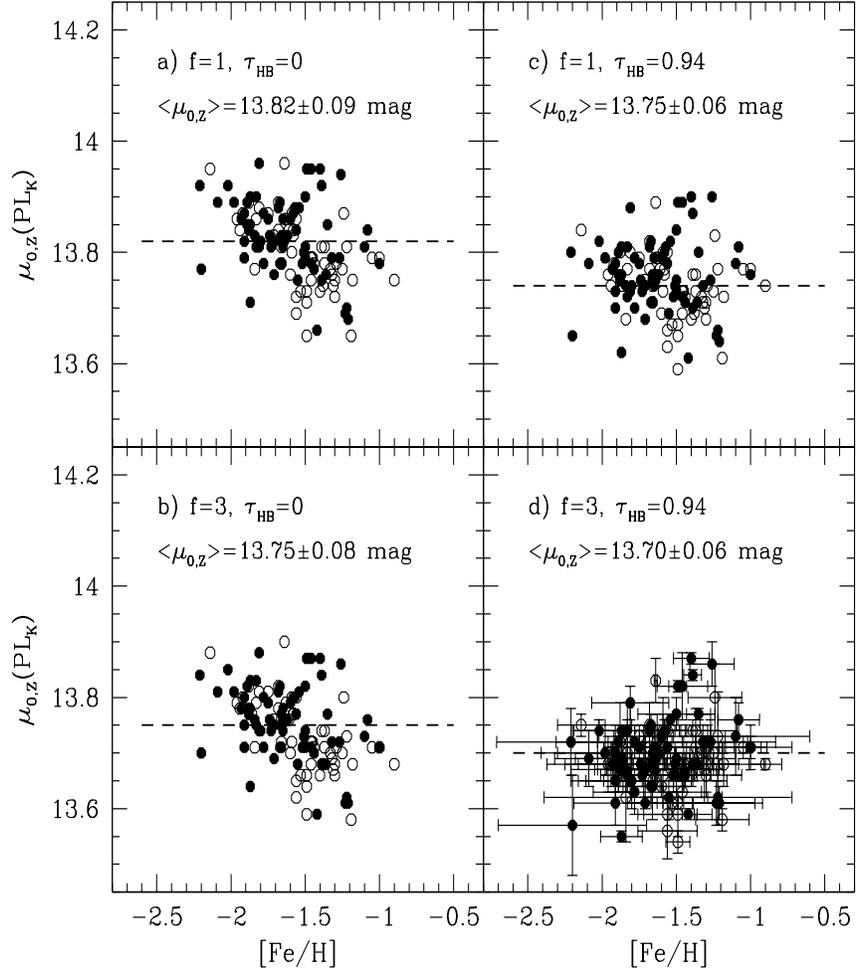}
\caption{Individual true distance moduli versus [Fe/H] for RR Lyrae in \omcp.
The panels show distance evaluations for scaled solar ($f$=1) and 
$\alpha$-enhanced ($f$=3) chemical compositions, as well as with HB-type 
$\tHB=0$ and 0.94. The error bars in the bottom right panel display 
individual errors affecting $K$ magnitudes and [Fe/H] abundances. 
The error bars in the other panels have been omitted to make more 
clear the distribution of F and FO variables.
\label{fig5}  
}
\end{center}
\end{figure}

We are now in a position to derive distance moduli of individual
RR Lyrae stars with metallicity measurements from Rey et al. (2000). 
In panel (d) in Fig. \ref{fig5} we show the resulting values as a
function of metallicity, leading to $\langle \mu_0 \rangle =13.70\pm 0.06$~mag.  
To illustrate the sensitivity of the method to
possible errors in the assumed HB type or $\alpha$-enhancement, we have
also computed the distance moduli for $\tHB=0$ and/or $f=1$ and plotted the 
results in Fig. \ref{fig5}. We note that the effect on the distance moduli is 
rather small considering that the estimated error on the HB type, $\sigma_\tau=0.2$
% point Q 
 (see star counts of HB stars in NGC~2808 listed in Table 2 of
Castellani et al. 2006), 
and on the $\alpha$-enhancement factor, $\sigma_f=0.5$, are typically smaller than 
the values adopted here. Of particular interest is the fact that the effect 
of an error in the assumed HB-type is to introduce a slope in these diagrams 
(see panels a and b).
According to the relations given in Table \ref{tab.logPM}, this trend 
can be explained as an empirical evidence that the more metal-poor RR Lyrae 
are connected with bluer HB populations when compared with the more 
metal-rich ones. 

\begin{figure}
 \begin{center}
\includegraphics[height=0.65\textheight,width=0.70\textwidth]{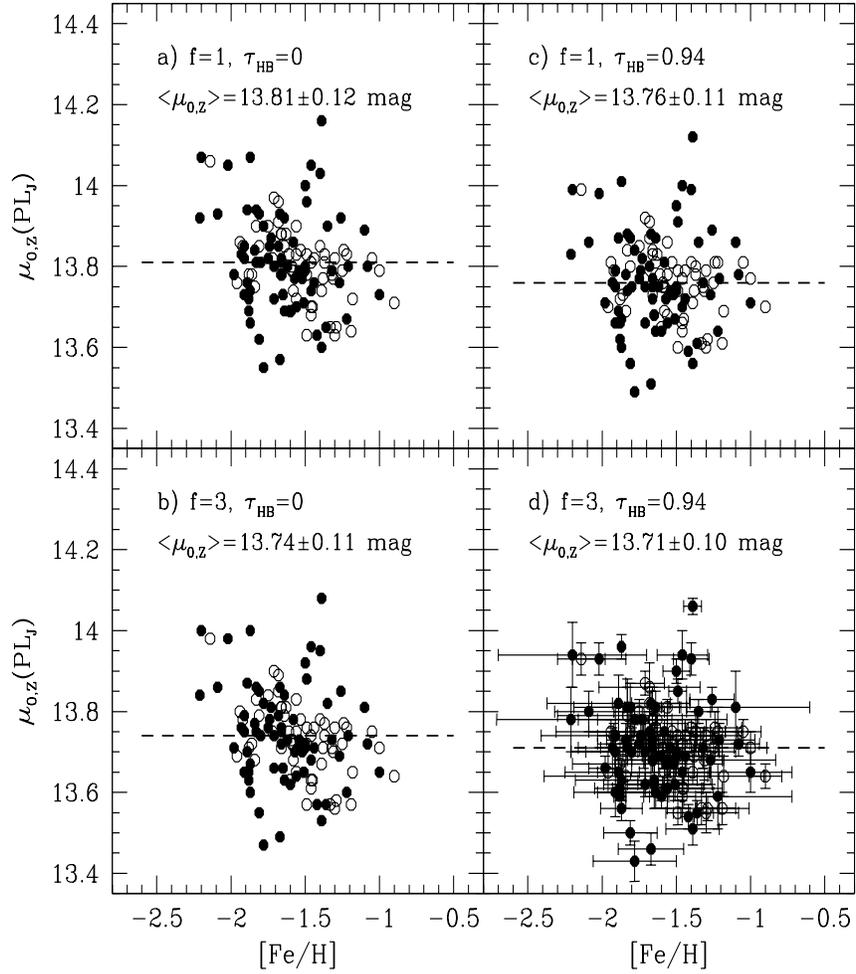}
\caption{Same as in Fig. \ref{fig5}, but for the mean $J$-magnitudes.  
\label{fig6}
 }
 \end{center}
 \end{figure}

The $K$-based distance determinations are fully supported by the
$J$-band measurements. The estimated uncertainty on the mean $J$-magnitudes
depends on the luminosity amplitude, ($\sigma_J(\rrc) \approx
0.1$; $\sigma_J (\rrab) \approx 0.2$), and it is less accurate than
for the $K$-band measurements ($\sigma_K \approx 0.015$).  
% point T 
 It is worth mentioning that we still lack empirical template curves for
the $J-band$, and therefore accurate mean $J-band$ magnitudes are only
available for 56 out of the 180 RR Lyrae in our sample.
As shown in Fig. \ref{fig6}, the  true distance 
moduli based on the relations given in Table \ref{tab.logPM} 
agree with the values based on $K$-magnitudes, but with a larger 
error.
% point T 
 Although, current mean $J-band$ magnitudes present this drawback, this
NIR band appears very promising to estimate distances, since it presents
several advantages over the $K-band$ and requires much less telescope
time to reach similar photometric accuracy.

We can also determine the distance to the individual RR Lyrae stars
using the \PLZk relations of Bono et al. (2003). By using both field 
and cluster RR Lyrae stars they found 
$M_K = -0.77 - 2.101 \log P_F + 0.231 \mbox{[Fe/H]}$
where the fundamentalized logarithmic period was computed 
as $\log P_F = \log P_{FO} + 0.127$.
Combining this relation with the sub-sample of RR Lyrae for which are 
available metallicity estimates from Rey et al. (2000), we find 
$\mu_0 = 13.77\pm0.07$ mag based on a total of 113 \rrab and \rrc stars.

In addition to the quoted intrinsic errors we estimate the external,
systematic error to be $\sigma_{\mbox{\scriptsize sys}} = 0.06$~mag.
This is the result of
summing in quadrature the contributions from the photometric zero point
($\sigma = 0.02$~mag), the uncertainty on the absorption $A_K$ ($\sigma =
0.01$~mag, based on uncertainties of $0.1$ on $R_K$ and $0.02$~mag on $E(B-V)$). 
% point W
 The uncertainty of the reddening is based on the recent Str{\"o}mgren photometric
investigation by Calamida et al. (2006), while the uncertainty on the selective
absorption coefficient is due to current uncertainties in the adopted extinction
law (Rieke \& Lebofsky 1985; Cardelli et al. 1989, and references therein). 
The absolute zero point of the
metallicity scale is a significant source of uncertainty as discussed by Dall'Ora
et al. (2004) due to systematic differences between metallicity scales. They
estimate $\sigma = 0.25$~dex on the zero point which results in the dominating
contribution to the systematic error of $\sigma = 0.25 \times 0.231 = 0.06$~mag. 

\begin{figure}
\begin{center}
\includegraphics[height=0.40\textheight]{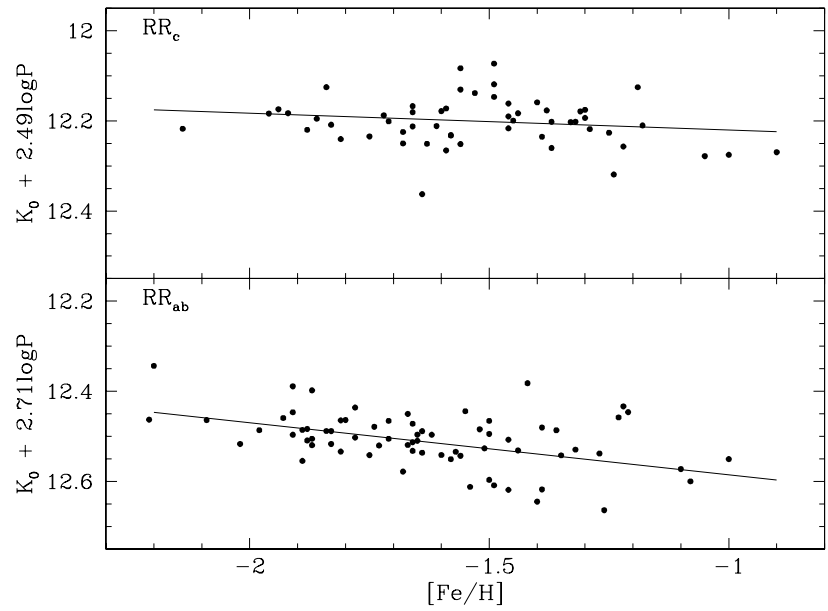}
\caption{The mean $K_0$ magnitude corrected for the empirically
determined period effect as a function of metallicity for fundamental
(\rrab) and first-overtone (\rrc) pulsators respectively. The best fit
lines have been overplotted.\label{fig7}
} 
\end{center}
\end{figure}

As the \omc RR Lyrae stars
exhibits a broad range of metallicities and all belong to
the same cluster, we can empirically constrain the metallicity effect
without having to rely on uncertain distance estimates. We
have made linear three-parameter fits of the relation in Eq.\ref{eq.PLZkFemp} for
\rrab and \rrc stars separately, using only dereddened {\em apparent} magnitudes, 
and using only the variables with individual $hk$ abundance evaluations. We find

\begin{eqnarray}
\label{eq.PLZkFemp}
K_0 (F)  & = & 12.70(\pm0.06) - 2.71(\pm0.12) \log P_F    + 0.12(\pm0.04) \mbox{[Fe/H]}  \hspace{0.7truecm} (N=54)\\
\label{eq.PLZkFOemp}
K_0 (FO) & = & 12.26(\pm0.05) - 2.49(\pm0.15)\log P_{FO} + 0.04(\pm0.04) \mbox{[Fe/H]}   \hspace{0.5truecm}  (N=48) \\ 
\label{eq.PLZjFemp}
J_0 (F)  & = & 13.09(\pm0.12) - 2.37(\pm0.25)\log P_F    + 0.15(\pm0.08)\mbox{[Fe/H]}    \hspace{0.7truecm}  (N=54)  \\ 
\label{eq.PLZjFOemp}
J_0 (FO) & = & 12.58(\pm0.08) - 2.09(\pm0.21) \log P_{FO} - 0.01(\pm0.05)\mbox{[Fe/H]}  \hspace{0.5truecm} (N=48) 
\end{eqnarray}

where the symbols have their usual meaning. In Fig. \ref{fig7} we have plotted 
the observed magnitudes
against metallicity after having removed the period dependency. The
metallicity effect is weak for the fundamental pulsators and for the
first-overtone pulsators the metallicity effect is consistent with 
only a marginal dependence.  In both cases the relations are in reasonable 
agreement with the predictions from the SHB models.
%point V 
 Current $K-band$ PL relations present two substantial differences when
compared with the PLZ relation derived by Bono et al. (2003):
a) we derived independent PL relations for F and FO pulsators, while Bono
et al. only derived semi-empirical "average" PL relations for F and
fundamentalized FO pulsators;
b) current PL relations only apply to cluster variables for which we can
estimate the HB type, while the PLZ relation derived by Bono et al. (2003)
applies to individual RR Lyrae stars. 

%%%%%%%%%%%%%%%%%%%%%%%%%%%%%%%%%%%%%%%%%%%%%%%%%%%%%%%%%%%%%%%%%%%%%%%%%%%%%
\section{Conclusions and final remarks}\label{Distance_01}

In Table \ref{tab.distances} we have listed recent empirical distance
estimates to \omc where we have corrected all the moduli for the same
reddening and employed the same reddening law as we have used for the
RR Lyrae stars. We note that our new near-IR results are in excellent 
agreement with the largely geometric distance to the eclipsing binary OGLEGC-17
from Thompson et al. (2001) and Kaluzny et al. (2002), and from the First 
Overtone Blue Edge method based on pulsation predictions from Caputo et al. (2002).
%point T 
 Current distance determinations based on $J$ and $K-band$ measurements
and on NIR PL($\tHB$) relations are also in very good agreement with
similar estimates based on NIR relations derived by CPS04. In particular,
for $\tau_{HB}=0.94$ and $f=3$ we found $13.70\pm0.06$ vs $13.72\pm0.06$
($K-band$) and $13.71\pm0.10$ vs $13.76\pm0.10$ ($J-band$). Note that
distance estimates based on CPS04 relations were determined using the
same solar metal abundance and we did not include a few very metal-poor
RR Lyrae, since CPS04 prediction range from Z=0.0005 to Z=0.006.
This finding is also supported by most recent $M_V-$[Fe/H] relations. By adopting the 
mean V-magnitudes for RR Lyrae stars provided by Kaluzny et al. (2004) and by 
Butler et al. (1978) together with the calibration given by Bono et al. (2003) 
we obtain for RR Lyrae with individual metal abundances a 
distance value of $(m-M)_0 = 13.72\pm0.11$. 
% point Y 
 We obtain a similar distance modulus ($13.62\pm 0.11$) using the same magnitudes
and the recent calibration of the $M_V-$[Fe/H] relation provided by Catelan (2005).
The slope of such a relation was estimated by Cacciari \& Clementini (2003), while
the zero-point is based on the trigonometric parallax of RR Lyr itself (Benedict et
al. 2002; Bono et al. 2002). 
Note that the systematic error for distance estimates based on visual magnitudes 
is $\sigma=0.08$, due to the significant contribution from the absorption 
correction.

On the other hand, we deviate by more than three $\sigma$ from the distance 
evaluation based on High Amplitude $\delta$ Scuti (HADS) stars 
($\mu_0=14.05\pm0.02$, McNamara 2000) and from the modulus of 
$\mu_0 = 13.41\pm0.13$ from van de Ven et al. (2006) based on 
cluster star dynamics.  It is worth noting that current uncertainties on 
metal abundance and on NIR photometry can not account for such a discrepancy. 

The NIR $PL$ relations for RR Lyrae stars provide an accurate 
approach for determining absolute distances to globular clusters
and other stellar systems. We find for \omc a distance modulus of
$13.70\pm0.06\pm0.06$ where the error estimates are random and
systematic errors, respectively. We note that for populous systems 
like globulars and dwarf galaxies the dominating source of error 
are the systematic sources. The accuracy of NIR photometry in crowded 
regions significantly improved during the last few years and
the sample sizes will certainly benefit from the new large field
of view NIR cameras available 
at 4m class telescopes (van der Bliek et al. 2004). New $J-$band light 
curve templates for both fundamental and first overtone RR Lyrae 
are also urgently required to improve the accuracy of their mean 
magnitudes in this band.

%%%%%%%%%%%%%%%%%%%
% Acknowledgments %
%%%%%%%%%%%%%%%%%%%

\acknowledgements
%During the revision of this manuscript Vittorio Castellani passed away
%on May 19, 2006. His suggestions, ideas, and personality will be 
%greatly missed.\\ 
It is a pleasure to thank S. Cassisi and A. Pietrinferni for several 
helpful discussions and for providing detailed sets of HB models. 
We also thank M. Marconi for many enlightening suggestions concerning 
the pulsation properties of RR Lyrae stars. We acknowledge an anonymous
referee for his/her positive comments and insights on an early version  
of this paper. 
We are very grateful to the ESO support astronomers in La Silla and 
in Garching for their ongoing support and effort in dealing with 
series of different Observing Blocks. We also thank the ESO Science 
Archive for their prompt support.  
This work was partially supported by Particle Physics and Astronomy 
Research Council (PPARC) and INAF/PRIN2005. 
This publication makes use of data products from the Two Micron All Sky
Survey, which is a joint project of the University of Massachusetts and
the Infrared Processing and Analysis Center/California Institute of
Technology, funded by the National Aeronautics and Space Administration
and the National Science Foundation.

%%%%%%%%%%%%%%
% References %
%%%%%%%%%%%%%%

%\pagebreak 

\clearpage

\begin{deluxetable}{llccc}
\tabletypesize{\scriptsize}
\tablecaption{Predicted near-infrared Period-Luminosity relations.
The pulsation modes are indicated as F for fundamental and FO for 
first-overtone pulsators.\label{tab.logPM}}
\tablewidth{0pt}
\tablehead{
\colhead{$Z$}  &
\colhead{mode}  &
\colhead{$a$}  &
\colhead{$b$}  &
\colhead{$c$}  
}
\startdata
\multicolumn{5}{c}{$M_K=a+b $log$P$+$c \tHB$}\\
0.0001 &F  & $-$1.32$\pm$0.02 & $-$2.19$\pm$0.01 & 0.14$\pm$0.01\\
0.0001 &FO & $-$1.65$\pm$0.02 & $-$2.32$\pm$0.01 & 0.14$\pm$0.01\\
0.0003 &F  & $-$1.23$\pm$0.02 & $-$2.34$\pm$0.01 & 0.04$\pm$0.01\\ 
0.0003 &FO & $-$1.55$\pm$0.02 & $-$2.38$\pm$0.01 & 0.04$\pm$0.01\\
0.001  &F  & $-$1.14$\pm$0.01 & $-$2.37$\pm$0.01 & 0.01$\pm$0.01\\
0.001  &FO & $-$1.48$\pm$0.01 & $-$2.41$\pm$0.01 & 0.01$\pm$0.01\\
0.004  &F  & $-$1.05$\pm$0.01 & $-$2.46$\pm$0.01 & 0.0\\
0.004  &FO & $-$1.37$\pm$0.01 & $-$2.43$\pm$0.01 & 0.0\\

\multicolumn{5}{c}{$M_J=a+b $log$P$+$c \tHB$}\\

0.0001 &F  & $-$0.92$\pm$0.02 & $-$1.62$\pm$0.01 & 0.08$\pm$0.01\\
0.0001 &FO & $-$1.13$\pm$0.02 & $-$1.64$\pm$0.01 & 0.07$\pm$0.01\\
0.0003 &F  & $-$0.83$\pm$0.02 & $-$1.77$\pm$0.01 & 0.05$\pm$0.01\\
0.0003 &FO & $-$1.10$\pm$0.02 & $-$1.83$\pm$0.01 & 0.04$\pm$0.01\\
0.001  &F  & $-$0.76$\pm$0.01 & $-$1.89$\pm$0.01 & 0.01$\pm$0.01\\
0.001  &FO & $-$1.04$\pm$0.01 & $-$1.88$\pm$0.01 & 0.01$\pm$0.01\\
0.004  &F  & $-$0.70$\pm$0.01 & $-$2.13$\pm$0.01 & 0.0\\
0.004  &FO & $-$0.94$\pm$0.01 & $-$1.93$\pm$0.01 & 0.0\\
\enddata
%\tablenotetext{a}{Metal abundance.}
%\tablenotetext{b}{Pulsation mode: fundamental (F), first-overtone (FO).}
%\tablenotetext{c}{Coefficients of the pulsation relations.}
\end{deluxetable}

\clearpage

\begin{deluxetable}{lll}
\tabletypesize{\scriptsize}
\tablecaption{Distance determinations to \omcp. \label{tab.distances}}
\tablewidth{0pt}
\tablehead{
\colhead{Method} &
\colhead{$(m-M)_0$\tablenotemark{a}} &
colhead{Notes}
}
\startdata
PL-$\tHB$ ($K$)       & $13.70\pm0.06$ & ~~~~~~~~~~~~~~\tablenotemark{b} \\
PL-$\tHB$ ($J$)       & $13.71\pm0.10$ & ~~~~~~~~~~~~~~\tablenotemark{b} \\
\PLZk                 & $13.77\pm0.07$ & ~~~~~~~~~~~~~~\tablenotemark{c} \\ 
PL-$\tHB$ ($K$)       & $13.72\pm0.06$ & ~~~~~~~~~~~~~~\tablenotemark{d} \\
PL-$\tHB$ ($J$)       & $13.76\pm0.10$ & ~~~~~~~~~~~~~~\tablenotemark{d} \\
$M_V-\mbox{[Fe/H]}$   & $13.72\pm0.11$ & ~~~~~~~~~~~~~~\tablenotemark{e} \\
$M_V-\mbox{[Fe/H]}$   & $13.62\pm0.11$ & Catelan et al. (2005)\tablenotemark{f} \\\\
Eclipsing binary      & $13.71\pm0.11$ & Thompson et al. (2001)\\
Eclipsing binary      & $13.75\pm0.04$ & Kaluzny et al. (2002)\\
FOBE                  & $13.74\pm0.11$ & Caputo et al. (2002)\\
High ampl. $\delta$ Sct& $14.05\pm0.02$& McNamara (2000)\\
Cluster dynamics      & $13.41\pm0.13$ & van de Ven et al. (2006)\\
\enddata
\tablenotetext{a}{Using the same mean reddening ($E(B-V)=0.11\pm0.02$) and 
reddening law (Cardelli et al. 1989) adopted in this paper. The error 
estimates do not include the contribution of systematic errors 
($\sigma_{sys}^K = 0.06$~mag, $\sigma_{sys}^V = 0.08$~mag).}
\tablenotetext{b}{Based on the PL-$\tHB$ relations presented here for 
$\tHB=0.94$, an $\alpha$-element enhancement factor $f=3$, and the new 
solar metal abundance.}
\tablenotetext{c}{Based on the semi-empirical $K-$band PLZ relation from 
Bono et al. (2003).}
\tablenotetext{d}{Based on the PL-$\tHB$ relations from Catelan et al.
(2004) and $\tHB=0.934$, an $\alpha$-element enhancement factor $f=3$, 
and the new solar metal abundance.}
\tablenotetext{e}{Based on the $M_V-$[Fe/H] relation from Bono et al.  (2003) 
and mean V-magnitudes from Kaluzny et al. (2004) and Butler et al. (1978).}
\tablenotetext{f}{ Based on the $M_V-$[Fe/H] relation from Catelan (2005) and 
mean V-magnitudes from Kaluzny et al. (2004) and Butler et al. (1978).}
\end{deluxetable}

\clearpage

\clearpage

\clearpage

\clearpage

\clearpage

\clearpage

\clearpage

\end{document}